\documentclass[12pt,fleqn]{article}
\input{source}
\begin{document}

\date{~} 

\title{Autoregressive models for time series of random sums of positive variables: application to tree growth as a function of climate and insect outbreaks} 

\author[a]{Zinsou Max Debaly,}
\author[b]{Philippe Marchand}
\author[b,c,d]{and Miguel Montoro Girona}
\affiliation[a]{CREST-ENSAI, UMR CNRS 9194,\\
Campus de Ker-Lann, 51 rue Blaise Pascal, BP 37203, 35172 Bruz cedex, France}
\affiliation[b]{INSTITUT DE RECHERCHE SUR LES FORÊTS,\\
445, boul. de l'Université
Rouyn-Noranda QC  J9X 5E4 ,   Canada.}
\affiliation[c]{GROUPE DE RECHERCHE EN ECOLOGIE DE LA   MRC-Abitibi (GREMA)
341, rue Principale Nord, Amos, QC, J9T 2L8, Canada13}
\affiliation[d]{RESTORATION ECOLOGY RESEARCH GROUP, Department of Wildlife, Fish and Environmental Studies, Swedish University of Agricultural Sciences, Skogsmarksgränd, Umeå 907 36, Sweden}
\emailAdd{zinsou-max.debaly@ensai.fr}
\emailAdd{philippe.marchand@uqat.ca}
\emailAdd{miguel.montoro@uqat.ca}

\maketitle

\onehalfspacing

\vspace{1cm}
\begin{abs}
We present a broad class of semi-parametric models for time series of random sums of positive variables. Our methodology allows the number of terms inside the sum to be time-varying and is therefore well suited to many examples encountered in the natural sciences. We study the stability properties of the models and provide a valid statistical inference procedure to estimate the model parameters. It is shown that the proposed quasi-maximum likelihood estimator is consistent and asymptotically Gaussian distributed. This work is complemented by simulation results and applied to  time series representing growth rates of white spruce (\textit{Picea glauca}) trees from a few dozen sites in Québec (Canada). This time series spans 41 years, including one major spruce budworm (\textit{Choristoneura fumiferana}) outbreak between 1968 and 1991. We found significant growth reductions related to budworm-induced defoliation up to two years post-outbreak. Our results also revealed the positive effects of maximum summer temperature, precipitation, and the climate moisture index on white spruce growth. We also identified the negative effects of the climate moisture index in the spring and the maximum temperature of the previous summer. However, the model's performance on this data set was not improved when the interactions between climate and defoliation on growth were considered. This study represents a major advance in our understanding of budworm--climate--tree interactions and provides a useful tool to project the combined effects of climate and insect defoliation on tree growth in a context of greater frequency and severity of outbreaks coupled with the anticipated increases in temperature. \end{abs}
 

\vspace{2cm}
\section{Introduction}
Many ecological studies require measuring the positive dependent variables of random numbers of statistical individuals sampled over time \citep{montoro2019post}. This approach is often necessary, as 1) researchers cannot observe the entire population, and 2) the individuals observed by researchers depend on time-varying resources. Applications of this statistical approach include studies of species behaviour and ecological services. In forestry, for example,  we can be interested in time series that represent the mass or size of a given tree species. We then randomly sample individual trees each year and observe the corresponding mass or volume, e.g., see \citet{VOURLITIS2022119917}. This approach is also applied to evaluate the area occupied by a species in relation to the available resources over time \citep{labrecque2020dynamics}. In fisheries, we can use this approach to track temporal changes in the weight of fish caught, e.g., \citet{w12112974}. 

In this paper, we evaluate the impact of climate change and insect outbreak on tree growth as recorded by growth rings. Spruce budworm (\textit{Choristoneura fumiferana}; SBW) is the most important defoliator of conifer trees in the eastern North American boreal forest \citep{montoro2018secret}. In the province of Québec (Canada), the forest area affected by this species of Lepidoptera over the last century covers more twice the size of the Ukraine \citep{navarro2018changes}. At the epidemic stage, massive populations of larvae cause widespread damage to tree foliage \citep{lavoie2019vulnerability}. SBW affects the main conifer boreal species in Canada, including balsam fir (\textit{Abies balsamea}), white spruce (\textit{Picea glauca}), and black spruce (\textit{Picea mariana}), resulting in a major impact on boreal forest regeneration and dynamics \citep{martin2020driving}. Moreover, SBW outbreaks produce important economic consequences through the loss of forest productivity. 

Previous works have studied the changes of forest composition following insect outbreaks, e.g., \citet{MORIN2021463}, the response of SBW outbreaks to climate change, e.g, \citet{fleming1995effects} and \citet{10.3389/fevo.2020.544088}, and demography, i.e., the rate of mortality of spruce during outbreaks \citep{gauthier2015boreal}. However, despite the major implications of future climate change, we continue to have a limited understanding of the combined effects of SBW outbreak and climate change on tree growth. Given that temperature and its variations as well as the timing and amount of precipitation affect organisms' survival, reproduction cycles, and spatial dispersion \citep{aber2001forest}, it is critical to understand the links between SBW outbreaks, climate, and tree growth to improve our understanding of the impacts of  future climate change on forest productivity \citep{klapwijk2013forest}. This concern is amplified by the expected increase in SBW outbreak severity and frequency under future climate scenarios \citep{navarro2018changes, seidl2017forest}.
 
In this paper, we contribute to filling this gap by proposing a broad class of semi-parametric models for positive-valued time series. Time-series data are common in forestry and the standard statistical approaches include descriptive exploratory techniques and linear mixed-effect models with time-varying variables on transformed data, e.g., \citet{montoro2016radial} and \citet{boulanger2004spruce}, and correlated error terms \citep{GirardinE8406}. However, these approaches suffer from several drawbacks. Descriptive exploratory techniques do not allow the drawing of inferences from the data, and linear mixed-effect models, as shown in several papers, e.g., \citet{chou2015range}, have demonstrated that specifying a linear model on transformed data often leads to a poor predictive performance. Applying a log transformation, for instance, can make the positive-valued data more normal; nonetheless, the obtained predicted value underestimates the expected value because of Jensen's inequality. Furthermore,  models having autocorrelated error terms do not account for the complex, dependent structure of tree-ring growth. To provide a more robust and reliable approach, we present a class of semi-parametric autoregressive models and use them to investigate the relationships between climate, SBW outbreak, and the growth of white spruce. We also discuss the advantages of applying a repeated-measures design.

Many previous studies have focused on modelling non-Gaussian time series, such as positive-valued processes. Gaussian processes can be represented as linear models, whereas time series of count or binary data are modelled by non-linear dynamics, e.g., \citet{sim1990first},  \citet{weiss2018introduction}, and \citet{davis2016handbook} and references therein. For positive-valued time series data, \citet{10.2307/2999632} proposed the range volatility model as an alternative to \emph{GARCH} models in finance, and its use has  rapidly expanded because of its diverse applicability. We refer the interested reader to the review by \citet{chou2015range}. Recently, \citet{aknouche_francq_2020} considered a positive-valued time series whose conditional distribution had a time-varying mean dependent on exogenous variables. Our approach here differs slightly from theirs, as the positive process under consideration is itself the sum of a random number of other positive variables.
 
Our approach is driven strongly by the available data at hand, which consist of multiple time series collected from several sites, where the number of sampled individuals varies over time and between sites. Hence, considering an aggregate value such as the sum or the mean of growth rings can produce a loss of variability linked to the sampling scheme. Moreover, in fields such as finance, some modelling relies on considering empirical quantities such as realized volatility; historical returns of investment products within a defined period are then analysed, e.g., \citet{allen2010realized}. Unlike our framework, which is typical for ecological studies, all transactions on investment products are recorded, i.e., the entire statistical population is observed. 

Our paper is organized as follows. In Section \ref{sec::definition}, we define the model used throughout this paper and discuss our modelling choice. The time-series properties of the models are also assessed in this section. We then present the maximum-likelihood based inference and its asymptotic properties in Section \ref{sec::mle}. Section \ref{sec::empirical} contains a small simulation study and an application to empirical data related to the growth of white spruce. All auxiliary lemmas and mathematical proofs are presented in 
section \ref{sec::proofs}.

\section{Models and stability results}\label{sec::definition}
We introduce here a generalized linear dynamic model for time series of random sum of positive variables, motivated by the empirical application where we analyze the annual growth of spruce trees subject to climate variation and outbreaks of SBW. 
In this case, growth is measured by taking cores at 1.30 m heigh from the trunk of a sample of trees in a forest \citep{montoro2017understanding}.  The samples were prepared, measured and analyzed conforming to standard dendroecological protocol \citep{krause1995changes}. Cores were air-dried, mounted on wood boards and sanded before tree rings were measured with WinDendro system \citep{guay1992new} or a manual Henson micrometer with an accuracy of 0.01 mm. The tree-ring series measurements covered the last 41 years, and were cross-dated using TSAP-Wi (Rinntech, Heidelberg, Germany).

We denote by $Y_{k,t}, t\in\Z, k = 1, \ldots, K$ the time series of the total basal area increment related to the $k-$th observational site, i.e. the sum of the increases in trunk cross-sectional area for the $n_{k,t}$ trees sampled for site $k$ on year $t$. We aim to model the dynamics of this process both in terms of its own past and $m$ additional covariates $X_{k,t}  \in \R^m$. In the empirical application presented in section \ref{sec::empirical}, the covariate process encompasses climate variables such as temperature and precipitation, as well as the level of defoliation due to SBW in previous years.  
Our model is given by : 

\begin{equation}
    \label{eq::modelLogLinearWithoutRandomEffect}
    Y_{k,t} = \sum_{l = 1}^{n_{k,t}} \zeta_{l, k,t}
\end{equation}
where conditionally on  $n_{k,t}, X_{k,t},n_{k,t}^- = (n_{k,t-s}, s \geq 1)$ and  $Y_{k,t}^- = (Y_{k,t-s}, s\geq1)$, the variables $\zeta_{l, k,t}, ~ 1\leq l \leq  n_{k,t},$ representing the basal area increments of individual sampled trees, are independent and identically distributed like a random variable $\zeta_{k,t}$ of mean $\lambda_{k,t}$. Moreover, $(n_{k,t})_{t\in\Z}$ is a sequence of \emph{i.i.d} random variables and conditionally on $n_{k,t}^-, $ the variable $n_{k,t}$ is independent from $X_{k,t}$ and $Y_{k,t}^-$. The mean process is given by
\begin{equation}
    \label{eq::lambda}
     \varphi_\delta(\lambda_{k,t}) =: \eta_{k,t} =  \omega_k + \sum_{j = 1}^p \alpha_j \frac{Y_{k, t-j}}{n_{k,t-j}} +    \beta^\top X_{k,t},~ \quad  k = 1,\ldots,K \text{ and }t = 1, \ldots, T,
\end{equation}
 $\omega_k  \in \R, \alpha_j \in \R,    \beta=(\beta_1, \ldots, \beta_m) \in \R^m$ and $\varphi_\delta$ is a real-valued function defined on $\R_+$ that can depend on a parameter $\delta.$ It is worth mentioning, without loss of generality, that the covariate process considered at time $t$ is included in the specification of $\lambda_{k,t}$ since multiple lags of a given set of variables can be included by simply stacking them into a vector. It is for example the case of the defoliation level in our application, since growth can be affected by defoliation up to 5 years prior (from $t-5$ to $t-1$).
 
  The variables $\zeta_{k,t}$ will be referred to as the unity random variables. We do not make any assumption about the distribution of the variables $\zeta_{k,t}$. Any distribution on $(0, +\infty)$ can be chosen. For example, an Exponential distribution with parameter $1/\lambda_{k,t}$,  log-Normal distribution with parameters $\log \lambda_{k,t} - \sigma^2/2$ and $\sigma$ or a Gamma distribution with parameters $\alpha \lambda_{k,t}$ and $\alpha$, to name a few.
 Whatever the distributions of unity random variables are, the conditional expectation of  $Y_{k,t}$  is $n_{k,t} \lambda_{k,t}$. However, under the assumption of the indenpendence of $\zeta_{l, k,t}, ~ 1\leq l \leq  n_{k,t},$ if they are exponentially distributed, the conditional variance is $n_{k,t}\lambda_{k,t}^2$  \emph{i.e} a  quadratic function of $\lambda_{k,t}$. For our example of Gamma-distributed unity random variables, the conditional variance is $n_{k,t}\lambda_{k,t}/\alpha$, \emph{i.e.} a linear function of $\lambda_{k,t}$. But in the case of the log-Normal distribution, the conditional variance is $n_{k,t}\lambda_{k,t}^4 (\exp\sigma^2 -1).$ With our semi-parametric framework, we will only focus on the estimation of regression parameters $\theta = (\delta, \omega_1,\ldots, \omega_K, \alpha_1, \ldots, \alpha_p, \beta^\top)^\top$ without the need to perform any distributional goodness of fit test.
 
 \begin{notes}{Copies of unity variables}
   In our general set up, the copies $\zeta_{l, k,t}, ~ 1\leq l \leq  n_{k,t}$ of the unity random variables $\zeta_{k,t}$ are not required to be independent. In practice where for example $\zeta_{l, k,t}$ represents the measure of annual growth for a sampled tree, the general assumption of identical distribution can be thought as a local stationary condition inside the site $k$ at time $t$.
  \end{notes}
 
 \begin{notes}{Marginal stationary distributions}
   Note from equations \eqref{eq::modelLogLinearWithoutRandomEffect}-\eqref{eq::lambda}, \\$Y_{k,t} = f_{\theta_k}(X_{k,t-s}, n_{k,t-s}, \zeta_{\ell,k,t-s}, s\geq 0, \ell\geq1)$ for {\small$\theta_k = (\delta, \omega_k, \alpha_1, \ldots, \alpha_p, \beta^\top)^\top.$} Then for $1\leq k_1\neq k_2 \leq K$ the distributions of $(Y_{k_1,0}, n_{k_1,0}, X_{k_1,0})$ and $(Y_{k_2,0}, n_{k_1,0}, X_{k_2,0})$ are not equal unless $\omega_{k_1} = \omega_{k_2} $ and $(X_{k_1,0}, \zeta_{\ell,k_1,0}, \ell \geq 1)$ is   equal in distribution to $(X_{k_2,0}, \zeta_{\ell,k_2,0}, \ell \geq 1)$. We will investigate in section \ref{sec::empirical} the consequences of the latter conditions on the proposed estimation procedure. 
 \end{notes}

 \begin{notes}{Form of regression function \eqref{eq::lambda}}
   Note that $\eta_{k,t}$ in \eqref{eq::lambda} does not depend linearly on $Y_{k,t-i}, i= 1, \ldots,p$, but on $Y_{k,t-i}/n_{k,t-i}, i= 1, \ldots,p.$ In fact, through \eqref{eq::lambda}, we make a link between the underlying mean process and the empirical estimate of the past mean process. Even for a constant size process, \emph{i.e.} $n_{k,t} = n_k, \forall t$, since the regression parameter $\alpha_i, i =1, \ldots,p$ is free of $k$, we still cannot yet express $\eta_{k,t}$ as a linear combination of $Y_{k,t-i}, i= 1, \ldots,p$. Moreover, one can expect $Y_{k,t-i}/n_{k,t-i} - \lambda_{k,t-i}, i= 1, \ldots,p$ or more generally $h(Y_{k,t-i}/n_{k,t-i}, \lambda_{k,t-i}), i= 1, \ldots,p$ for some mapping $h$ such that $\E h(Y_{k,t-i}/n_{k,t-i}, \lambda_{k,t-i}) = 0$ in \eqref{eq::lambda} at the place of  $Y_{k,t-i}/n_{k,t-i}, i= 1, \ldots,p.$ Indeed, with the latter two mentioned specifications, \eqref{eq::modelLogLinearWithoutRandomEffect}-\eqref{eq::lambda} define the so-called \emph{GLARMA} model (see for example, \cite{weiss2018introduction} for more details). In the  present form \eqref{eq::modelLogLinearWithoutRandomEffect}-\eqref{eq::lambda}  has some similarities with the well known \emph{ARCH} model (\citet{Bollerslev(1986)}). We leave the topic of  \emph{GLARMA} specification for furthers works.
 \end{notes}

 \begin{notes}{Contrast with the non-linear mixed model}
 The model \eqref{eq::modelLogLinearWithoutRandomEffect}-\eqref{eq::lambda} has some similarities with the well-known mixed models. Indeed, as for mixed models, the $\omega_k$ stands for the site fixed effect and the random effect in embedded is the distribution of unity variables. The simple example of 
 $\zeta_{l, k,t} = \lambda_{k,t}\epsilon_{l, k,t}$, where $(\epsilon_{l, k,t})_{l\geq1}$ is a sequence of identically distributed random variables of mean 1, fit with the so-called multiplicative form random effect models \citep{cameron2013regression}. But more complex random effects can be handled. However, the model \eqref{eq::modelLogLinearWithoutRandomEffect}-\eqref{eq::lambda} is more general since it allows the individuals sampled over time to change.  Indeed, as we will see in section \ref{sec::mle}, the individual measures are no longer needed when the sequence $(Y_{k,t}, n_{k,t})$  are available. Also, in terms of the application to resource management, it is often of interest to model and predict a population quantity like the sum of basal area growth in a forest.  
 \end{notes}

 \section*{Choice of the link function $\varphi$}
 
  The logarithmic link function is often applied and coincides with the well known log-linear model, see for example \citet{cameron2013regression} for models for count data. This link function assumes a linear relationship between the logarithm of the mean process and the covariates. However, there exist some other link functions that preserve the linear correlation at least on the positive part of $\R.$  Consider for example, the threshold mapping $x \mapsto \max(x,0)$. This mapping is not smooth and most of time, one makes some restrictions on model parameters to directly obtain the positiveness of the mean. Here, we will apply the inverse of the so-called softplus function as a link function. Indeed, the softplus function (see \citet{glorot2011deep}) is interesting for two reasons.
  The first one related to modelling is that it preserves the linearity on the positive part of real line. This is also pertinent for our biological application, as we expect a linear effect of covariates on growth above a certain threshold representing the minimal favorable conditions for growth. The minimum growth expected may not be exactly zero, which is why we will consider later a slightly different version of softplus that we will refer to as $\mathrm{softplus}_\delta$ for $\delta>0$ defined as $\mathrm{softplus}_\delta(x) = \log(1+\delta + \exp(x))$. The second one and technical advantage  is that the mapping $\mathrm{softplus}_\delta$ is infinitely differentiable. The Figure \ref{fig::softplusVsmax} in the Appendix shows the difference between the $\mathrm{softplus}_\delta$ link function and $\max(x,0)$ where $\mathrm{softplus}$ stands for $\mathrm{softplus}_0.$ One can note that $\mathrm{softplus}_\delta$ is lower bounded   by $\log(1+\delta).$

 \begin{notes}{Model Interpretation}
Obviously, with the softplus$_\delta$ link function, the mean process increases with the $j-$th covariate process if $\beta_j>0$ and decreases with this one when $\beta_j<0.$  Since softplus$_\delta(x) \sim_\infty x $, the mean process can be approximated, all other things remaining equal, by $\beta_j X_{j,t}$ for large values of $X_{j,t}$ and $\beta_j>0$ and then increases by $\beta_j \alpha$ for   increasing value $\alpha$ of $X_{j,t}$. Let us denote by RG$(x,y)$, the relative rate of growth of the mean process between $x$ and $y$ \emph{i.e} RG$_\delta(x,y) = \gamma_\delta(x)/\gamma_\delta(y)$ where $\gamma_\delta$ is the derivative function of softplus$_\delta$. For $\beta < 0,$  $\lim_{x\rightarrow\infty}$RG$_\delta(\beta (x+\alpha),\beta x) = \e^{\beta\alpha}.$ Therefore, the rate toward $\log(1+\delta)$ driven by $X_{j,t}$ is given by $\e^{\beta_j\alpha}$ when $\beta_j<0.$ Moreover, When $\delta \sim 0,$ by the l'Hôpital's rule $\lim_{x\rightarrow\infty}$RG$_0(x,y)  = \lim_{x\rightarrow\infty}$softplus$_0(\beta (x+\alpha))/$softplus$_0(\beta x)=\e^{\beta\alpha}.$ Therefore, all other things remaining equal, the mean process will be divided by  $\e^{\beta\alpha}$ when $X_{j,t}$  increases by $\alpha$ for large values of $X_{j,t}$ and $\beta_j<0.$
\end{notes}

  Theorem \ref{th::stability} gives some stability conditions of model \eqref{eq::modelLogLinearWithoutRandomEffect}-\eqref{eq::lambda} with the inverse of the softplus function as the link whereas Lemma \ref{lem::generalresult} in the section \ref{sec::proofs} stands for a general result for $\varphi$.

 \begin{theo}
 \label{th::stability}
 Under the assumptions \ref{ass::integrability}-\ref{ass::contraction} in   section \ref{sec::proofs} and $\sum_{j = 1}^p |\alpha_j| < 1$, there exists a unique set of $K$ stationary, ergodic sequences $(Y_{k,t},  n_{k,t}, X_{k,t}), k = 1, \ldots, K$ that are the solution of equations \eqref{eq::modelLogLinearWithoutRandomEffect}-\eqref{eq::lambda} with $\E|\eta_{k,0}|<\infty, k=1,\ldots,K$.
 \end{theo}

 \section{Estimation and asymptotics properties}\label{sec::mle}
This section is devoted to the estimation of the conditional mean parameters by the quasi-maximum likelihood estimator (QMLE) based on a member of the exponential family. We
consider the exponential QMLE (EQMLE) because this estimator coincides
with the maximum likelihood estimator (MLE) when the unity random variables follow the exponential $\Gamma(1,\lambda_{k,t}^{-1})$ distribution, and the copies $\zeta_{l, k,t}, ~ 1\leq l \leq  n_{k,t}$ are independent. 

For our application, the $K$ time series are observed between the time points $1$ and $T.$ We provide an asymptotic theory for the estimated parameters and present the results of a small simulation study investigating the finite-sample properties of the estimator. In the following section, we make $\lambda_{k,t}$ dependent on the parameter $\theta$($\in \Theta$ a compact set); that is 
$$\log(\exp\circ\lambda_{k,t}(\theta) - 1 - \delta) =  \omega_k + \sum_{j = 1}^p \alpha_j \frac{Y_{k, t-j}}{n_{k,t-j}} +    \beta^\top X_{k,t}=: \eta_{k,t}(\theta) ,~ \quad  k = 1,\ldots,K \text{ and }t = 1, \ldots, T ,$$ 
where $\delta \geq \delta\_ > 0$.
Let us denote the true, data-generating parameter value by $\theta_0$. 

The loss function from the exponential quasi-maximum likelihood is given by
\begin{equation}
\label{eq::loss}
    r_T(\theta) =  \sum_{k = 1}^K T^{-1} \sum_{t = 1}^T \left(\frac{Y_{k,t}}{\lambda_{k,t}(\theta)} + n_{k,t}\log \circ \lambda_{k,t}(\theta)\right) =:       \sum_{k = 1}^K T^{-1} \sum_{t = 1}^T \ell_{k,t} (\theta) =:  \sum_{k = 1}^K \ell_{k}(\theta)
\end{equation}
and 
\begin{equation}
\label{eq::estimatorQMLE}
    \hat \theta_T = \underset{\theta \in \Theta}{\mathrm{argmin}}~~r_T(\theta).
\end{equation}

The derivative of $\lambda_{k,t}(\theta)$ with respect to $\theta$ is given by
{
\begin{eqnarray*}
 \dpartial{\lambda_{k,t}(\theta)}{\theta} & =: & \dot\lambda_{k,t}(\theta) \\
 &= &\left(\frac{1}{1+\delta+\e^{\eta_{k,t}(\theta)}}, \frac{\e^{\eta_{k,t}(\theta)}}{1+\delta+\e^{\eta_{k,t}(\theta)}}\left(\iota_k, \frac{Y_{k,t-1}}{n_{k,t-1}}\ldots\frac{Y_{k,t-p}}{n_{k,t-p}},X_{k,t}^\top\right)\right)^\top.  
\end{eqnarray*}
}
where $\iota_k$ is a vector of size $K$ with $1$ at the $k-$th position and $0$ elsewhere.
We will denote by $\dot\lambda_{k,t}$ (resp. $\lambda_{k,t}$), the vector $\dot\lambda_{k,t}(\theta)$  (resp. $\lambda_{k,t}(\theta)$), evaluated at the point $\theta = \theta_0.$

We will study the asymptotic properties of the QMLE estimator (\ref{eq::estimatorQMLE}). To do so, we employ \citet{taniguchi2002asymptotic}  (Thm 3.2.23), which was extended in \citet{klimko1978conditional}. The lemmas in our Appendix produce the general result for the asymptotic properties of QMLE (\ref{eq::estimatorQMLE}). The following theorem represents the consistency and the asymptotic normality of (\ref{eq::estimatorQMLE}) for the $\mathrm{softplus}_\delta$ link function.
Let us set
$$
V_k = \E\left[\frac{1}{\lambda_{k,0}^{2}}\left(n_{k,0} - \frac{Y_{k,0}}{\lambda_{k,0}}\right)^2\dot\lambda_{k,0}\dot\lambda_{k,0}^\top \right] \text, { and }
J_k = \E\left[n_{k,0}\frac{1}{\lambda_{k,0}^{2}}\dot\lambda_{k,0}\dot\lambda_{k,0}^\top \right].
$$

\begin{theo}
\label{th::asymptotic}
Suppose that the assumptions \ref{ass::stabilityForTrueParameters}--\ref{ass::identifiability2} in section \ref{sec::proofs} are met. Then, almost surely,
$$
\lim_{T\to\infty} \hat \theta_T = \theta_0.
$$
If \ref{ass::independencySolution}--\ref{ass::uniformhessian} hold true and $\theta_0$ is located in the interior of $\Theta$,  
$$\lim_{T\to\infty}  \sqrt{T}(\hat \theta_T - \theta_0) = \mathcal{N}(0, J^{-1}VJ^{-\top}) ,$$
where $J = \sum_{k = 1}^K J_k$ and $V = \sum_{k = 1}^K V_k.$
\end{theo}

\section{Application}\label{sec::empirical}

\subsection{Simulation}

We examined the finite-sample performance of the QMLE presented in the previous section through a small simulation study. We present the results for QMLE under two different data-generating processes, here referred to as \emph{Scenario 1} and \emph{Scenario 2}, with $m=10$ covariates. First, $X_{k,t}$ does not depend on $k$ and is a sequence of \emph{i.i.d} random variables distributed as exponential random variables with means $\lambda_1, \ldots, \lambda_m.$ In the second, for a fixed $k,$ $X_{k,t}$ is sampled independently from exponential distributions of mean $0.4k\lambda_1, \ldots, 0.4k\lambda_m.$ For the two data-generating processes, for a fixed $k$, the process $(n_{k,t})_{t\geq1}$ is independently sampled from a Poisson distribution of mean $\tau_k$ as follows: 
for a fixed $K, \tau_1, \ldots, \tau_K$ are independent and distributed as an exponential random variable of mean $K.$ Moreover, we take $p=1$ and $\delta= 0.5, \beta = (0, 1, -1, 0.5, -0.5, -1.5, 1.5, -2, 2, 0), \alpha_1 = 0.6$, and $\omega_1, \ldots, \omega_K$ independently and uniformly sampling in the range $(-0.5K, 0.5K)$ for a fixed $K.$ We sequentially choose $K=5, 10, 15, 20$ and $T= 50, 100.$  The samples are nested, i.e., the sample for the first scenario and $K=5, T = 50$ is a subset of that of $K=5, T = 100$. Indeed, our aim here is to evaluate the consequences of increasing $K$ and $T$ on the performance of our estimator. For each sample, we compute the estimator (\ref{eq::estimatorQMLE}) and the corresponding theoretical standard errors (TSE) given by the Gaussian limit distribution. We replicate $B = 100$ times the experiment. Table \ref{tab:simulations} presents our simulation results. The line EQML refers to the average estimated value of the parameters, and TSE refers to the average value of estimated theoretical standard errors: 
$$
\mathrm{EQML} = B^{-1} \sum_{b = 1}^{B} \hat \theta_T^{(b)} , \text{ and } \mathrm{TSE} = B^{-1} \sum_{b = 1}^{B} \mathrm{diag}\left\{ {\hat J^{-1(b)}}  {\hat V}^{(b)} {\hat J^{-\top(b)}}\right\}^{1/2} ,
$$
where the superscript $b$ represents the index of replication, and $\mathrm{diag}M$  for a matrix $M$ is the diagonal elements of $M$. It appears that the model parameters are well estimated, except for $\omega_k, k = 1, \ldots, K$ when $K$ is very small relative to $T$, which coincides here with $K=5, T = 50, 100.$   We leave deep simulation studies for a future study.
 
\subsection{Application to the white spruce growth series}
Dendrochronology, i.e., the study of the time series of tree rings, is a powerful tool for reconstructing past natural and anthropic disturbances \citep{montoro2016radial}, \citep{boulanger2004spruce}, and \citep{labrecque2020dynamics}. Tree rings represent natural hard disks that record environmental changes and thereby offer the potential to understand the evolution of complex natural phenomena over time, such as disturbances. Dendrochronological data have provided a better understanding of insect outbreak dynamics \citep{navarro2018changes}, \citep{camarero2003impact}, and \citep{speer2017creating}.

Here we used the dendroecological series from \citet{jardon2003periodicite}, which includes annual tree-ring width measurements for 631 white spruce (\emph{Picea glauca}) trees distributed across 45 sites in southwestern Quebec, Canada, with 1 to 23 trees per site. These time series comprise between 63 and 247 rings. We converted the ring-width increments to basal-area increments (BAI) using the full series; however, because of covariate availability, we limited our analysis to the AD 1955--1995 period (41 years) to study only a single insect outbreak event (see Fig. \ref{fig::carteSites} in the Appendix).

We interpolated climate variables at the study sites for these 41 years  using BioSIM \citep{regniere2014biosim}, a software package that interpolates daily climate station data on the basis of latitudinal and elevational climate gradients and the spatial correlations estimated from 30-year climate normals. We computed the following climate summaries from daily data for the spring (April--June) and summer (July--September) seasons separately: mean of daily maximum temperatures, total precipitation, and the climate moisture index (CMI) equal to the difference between precipitation and potential evapotranspiration (PET). Daily PET values were estimated using the Penman--Monteith equation as implemented in the SPEI package \citep{pkgSPEI} in R, on the basis of BioSIM-interpolated values of the minimum and maximum temperature, wind speed at 2 m, solar radiation, dew point temperature, and atmospheric pressure, using the "tall" crop model in SPEI.

One major SBW outbreak occurred in Quebec during the study period, spanning from 1967 to 1991. We obtained annual estimates of the severity of the SBW outbreak at the location of each study site through defoliation maps produced by the Quebec Ministry of Forests, Wildlife and Parks \citep{mffp_tbe}. These maps are digitized versions of hand-drawn outlines of defoliated areas produced by aerial surveys of the affected regions. The defoliation level for each area is classified on a scale of 1 to 3 corresponding to a low (approx. 1\%--35\%), moderate (36\%--70\%) or high (71\%--100\%) fraction of the year's foliage defoliated by SBW. We note that these defoliation levels mainly reflect the status of balsam fir (\emph{Abies balsamea}), which is the main SBW host and is generally more severely affected than white spruce. Therefore, these defoliation levels are a proxy for outbreak severity, i.e., the potential herbivory pressure exerted by budworm on white spruce at the site.

Because tree growth and its vulnerability to both climate and defoliation depend on tree age, we split the data set and fit our models separately for the five age classes of $\leq$75, 75--100, 100--125, 125--150, and $\geq$150 years. We included as covariates the mean daily maximum of temperature, the total precipitation, and the mean CMI for the current and previous spring and summer. Only one of either precipitation and CMI appeared in a given model version because of the correlation between these two variables. We also included as covariates the defoliation levels for the five preceding years, a delay that estimates the time needed to fully regrow the lost foliage after an outbreak. Note that we do not expect defoliation to have a marked effect on the same year's growth ring \citep{krause2003temporal}. Finally, we considered models having interaction effects of the preceding year's defoliation level and climate variables, representing the possibility that climate conditions can increase or decrease a tree's sensitivity to SBW outbreaks.

Data processing and analyses were performed in R \citep{rcitation} with the package dplR \citep{bunn2008dendrochronology} used to process tree-ring data. We minimized the criterion (\ref{eq::loss}) with the R command \emph{nlm} (\citet{dennis1983numerical}). All developed software is available under the Creative Commons (CC) license (see data availability statements). We used the QAIC criterion for selecting the model. The primary analysis based on partial autocorrelation plots led us to select $p=1$.

\begin{figure}
    \centering
    \begin{tabular}{c}
     (a)\\
    \includegraphics[width = \textwidth, height = 0.25\textheight]{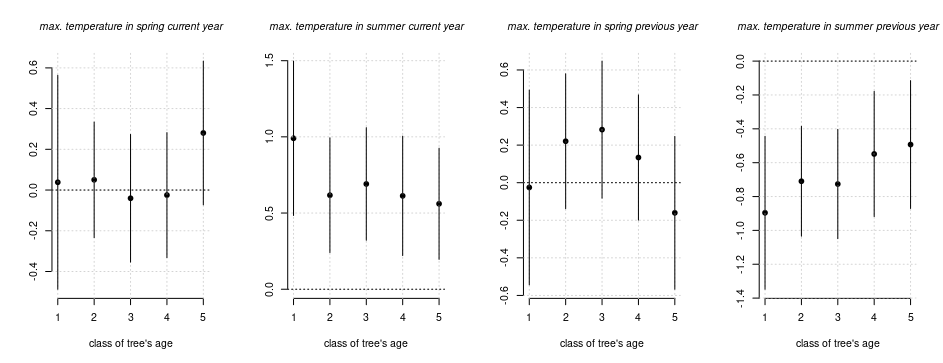}\\
    (b)\\
     \includegraphics[width = \textwidth, height = 0.25\textheight]{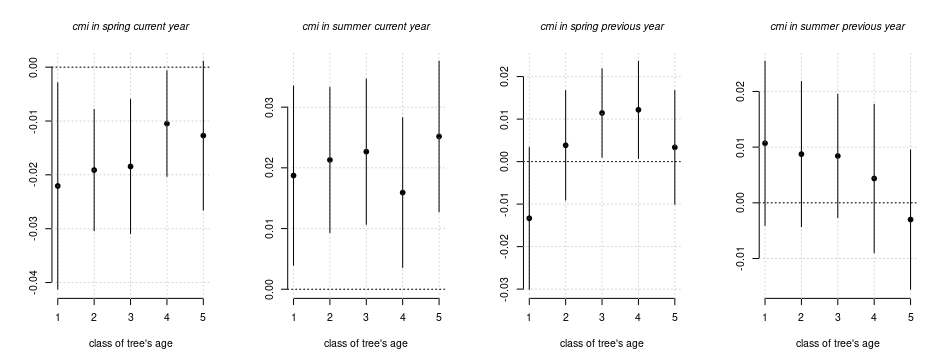}\\
    (c)\\
      \includegraphics[width = \textwidth, height = 0.25\textheight]{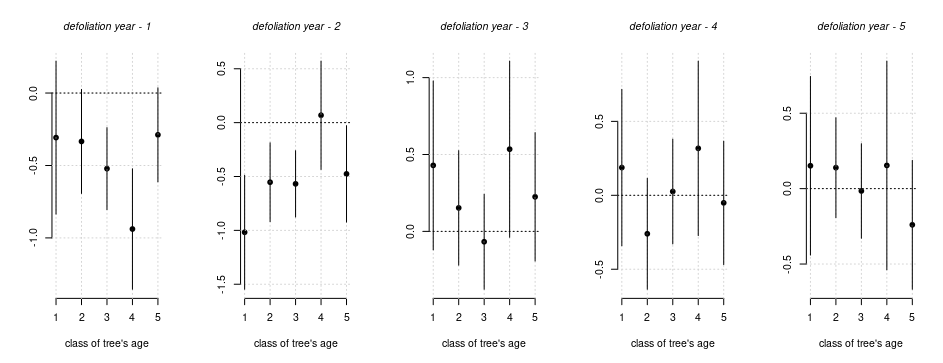}
      
    \end{tabular}
    \caption{Model with Temperature + CMI + Defoliation.  classes of age 1 : < 75, 2 : 75 -- 100, 3 :  100 -- 125, 4 : 125 -- 150 and 5 : > 150 years (a)  effects of maximum temperature in spring and summer in current  and previous year; (b)  effects of cmi index in spring and summer in current  and previous year and (c)  delayed effect of level of defoliation. The dashed horizontal line corresponds to zero.}
    \label{fig::modelwithcmi}
\end{figure}

\begin{figure}
    \centering
    \begin{tabular}{c}
     (a)\\
    \includegraphics[width = \textwidth, height = 0.25\textheight]{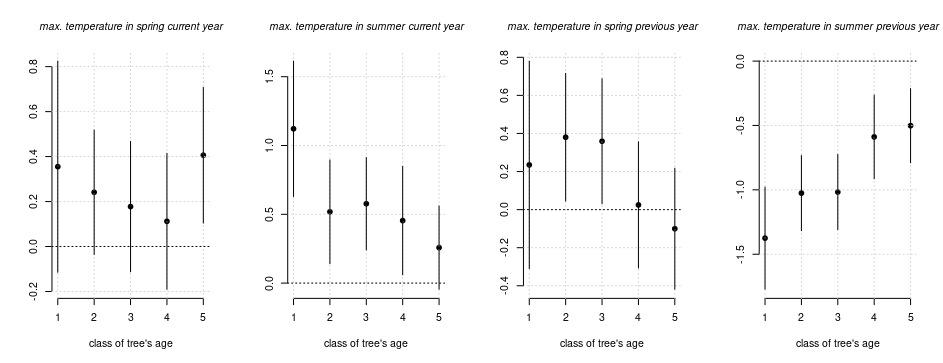}\\
    (b)\\
     \includegraphics[width = \textwidth, height = 0.25\textheight]{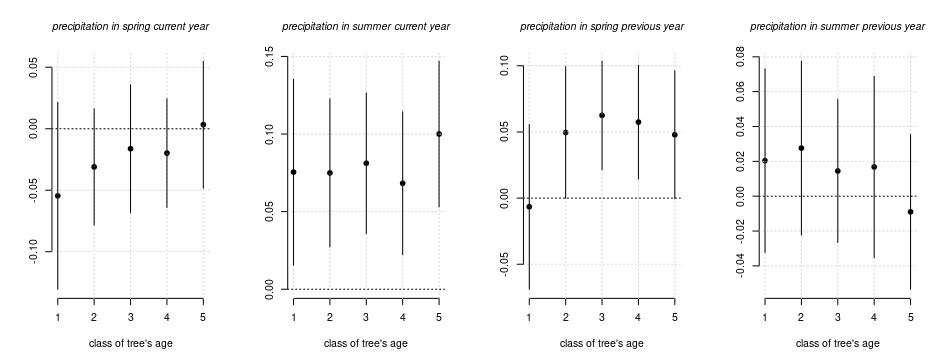}\\
    (c)\\
      \includegraphics[width = \textwidth, height = 0.25\textheight]{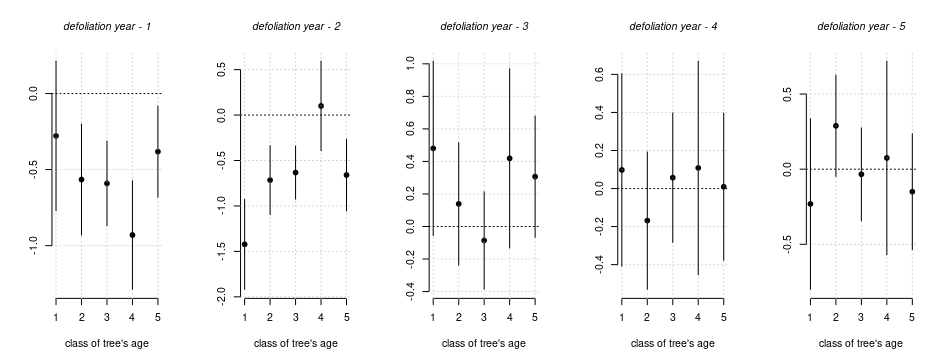}
    \end{tabular}
    \caption{Model with Temperature + Precipitation + Defoliation. classes of age 1 : < 75, 2 : 75 -- 100, 3 :  100 -- 125, 4 : 125 -- 150 and 5 : > 150 years (a)  effects of maximum temperature in spring and summer in current  and previous year; (b)  effects of precipitation index in spring and summer in current  and previous year and (c)  delayed effect of level of defoliation. The dashed horizontal line corresponds to zero.}
    \label{fig::modelwithprec}
\end{figure}

According to the QAIC, the best models were those lacking an interaction between climate and defoliation. Our model results (Figures \ref{fig::modelwithcmi} and \ref{fig::modelwithprec}) revealed that higher defoliation levels led to reduced tree-ring growth, but this effect vanished after two years; however, note that while the direct effect vanished, expected growth remained lower in the successive years because of the large estimated first-order autocorrelation coefficient (0.8--0.9, depending on age class). Moreover, there was no significant effect of defoliation on the following year's growth for the youngest and oldest trees although it produced an effect two years following the defoliation. The results differed markedly for middle-aged trees, which were significantly affected one year after defoliation but not in the second year. 

For the climate variables, high maximum temperatures in the summer increased growth, with up to a 5.6 cm\texttwosuperior\ increase in basal area from a 10 \textdegree C increase in summer maximum temperature. However, the previous summer's temperature had a negative effect on growth. Finally, the spring CMI was negatively correlated with tree-ring growth, whereas the summer CMI had a positive effect. However, both the CMI and precipitation in the previous spring increased the tree-ring growth of the current year: 100 mm greater precipitation led to at least a 6.8 cm\texttwosuperior\ increase in basal area growth.

\section{Proofs for the main results}\label{sec::proofs}
 Throughout this section, we will denote by $\zeta_{k,t}^\infty = (\zeta_{k,t,l})_{l\geq1}$, the   sequence of  copies of the unity random variables $\zeta_{k,t}.$    Moreover $\zeta_{k,t}$ can be decomposed into two components: its mean $\lambda_{k,t}$ function of $X_{k,t}$ and a free random variable  $\zeta_t$. For example, $\zeta_{k,t} = \lambda_{k,t}\zeta_t$ for a positive random variable $\zeta_t$ of mean 1.  We will write $\zeta_{k,t} := \zeta_{k,t}(\lambda_{k,t},  \zeta_t)$ to denote the relationship between $\zeta_{k,t} $ and $\lambda_{k,t}$ and $\zeta_t$. Accordingly, $\zeta_{k,t,l} = \lambda_{k,t}\zeta_{t,l}$ with $\zeta_{t,l}, l\geq1$ \emph{i.i.d} with a mean of 1 or in general $\zeta_{k,t,l} := \zeta_{k,t,l}(\lambda_{k,t},  \zeta_{t,l})$ with $\E\zeta_{k,t,l} = \lambda_{k,t}$. Let $\mathcal{F}_{k,t}$ denote the $\sigma$-algebra generated by $\zeta_{s}, X_{k,s+1}, s \leq t$ and $\mathcal{F}_{k,t,n}$  generated by $n_{k,s},\zeta_{s}, X_{k,s+1}, s \leq t.$ Finally, we will denote by $\phi_\delta$  the inverse of $\varphi_\delta : \phi_\delta(x) = \varphi_\delta^{-1}(x)$. For stability, we will consider the following set of assumptions: 
 
 \begin{enumerate}[label = (A.\arabic*)]
     \item \label{ass::generallink} The function $\phi_\delta$ is $\upsilon-$Lipschitz, and $\upsilon\sum_{i=1}^p |\alpha_i|<1.$
 \end{enumerate}
 
 \begin{enumerate} [label = (ST.\arabic*)]
     \item \label{ass::integrability}For $k = 1, \ldots, K, (n_{k,t-1}, \zeta_{k,t-1}^\infty, X_{k,t})_{t \in \Z}$ is stationary, ergodic, $(n_{k,t}, \zeta_{k,t}^\infty)$ is independent from $\mathcal{F}_{k, t-1,n}$, and $\E|X_{k, 0}|_1 < \infty.$
     \item \label{ass::contraction} For $k = 1, \ldots, K,$ 
     $$
     E(|\zeta_{k,t}(\lambda_{k,t},  \zeta_t) - \zeta_{k,t}(\overline\lambda_{k,t},  \zeta_t)| \vert \mathcal{F}_{k, t-1,n}) \leq |\lambda_{k,t} - \overline \lambda_{k,t}|.
     $$
 \end{enumerate}

 It is worth noting that the example $\zeta_{k,t} = \lambda_{k,t}\zeta_t$ for a positive random variable $\zeta_k$ of mean 1 verifies the condition \ref{ass::contraction}.

\begin{lem}
\label{lem::generalresult}
Under the assumptions \ref{ass::generallink}--\ref{ass::contraction}, there exists a unique set of $K$ stationary, ergodic sequences $(Y_{k,t}, n_{k,t}, X_{k,t}), k = 1, \ldots, K$ that are a solution of equations (\ref{eq::modelLogLinearWithoutRandomEffect}) and (\ref{eq::lambda}) with $\E|\eta_{k,0}|< \infty,  k=1,\ldots, K$.
\end{lem}
 The proof of Lemma \ref{lem::generalresult} uses the techniques of iterated random maps. We refer interested readers to \citet{debaly_truquet_2021} theorems 2 and 4, which investigated the problem of solving recursive stochastic equations with covariates or \citet{debaly2019stationarity} in the case where no covariates are included in the dynamic.
\paragraph{Proof of Lemma  \ref{lem::generalresult}}
From (\ref{eq::lambda}),
$$
\eta_{k,t} = \omega_k + \sum_{j = 1}^p \alpha_j \frac{1}{n_{k,t-j}}\sum_{\ell = 1}^{n_{k,t-j}}\zeta_{k,t-j,\ell}(\phi_\delta(\eta_{k,t-j}), \zeta_{t-j,\ell}) +    \beta^\top X_{k,t}.
$$
Then under the condition \ref{ass::integrability}, the processes $(\eta_{k,t} = \varphi_\delta(\lambda_{k,t})_{t\in\Z}, k = 1,\ldots,K$ obey some recursive stochastic equations,
 $$\eta_{k,t} = f(\eta_{k,t-1}, \ldots, \eta_{k,t-p}; n_{k,t-1},\ldots,n_{k,t-p},\zeta_{k,t-1}^\infty, \ldots, \zeta_{k,t-p}^\infty, X_{k,t}).$$
 And with \ref{ass::generallink}, for $k = 1,\ldots, K, (x,y)\in \R^{2p}$,
 {\footnotesize
 $$
 E(|f(x; n_{k,t-1},\ldots,n_{k,t-p},\zeta_{k,t-1}^\infty, \ldots, \zeta_{k,t-p}^\infty, X_{k,t}) - f(y; n_{k,t-1},\ldots,n_{k,t-p},\zeta_{k,t-1}^\infty, \ldots, \zeta_{k,t-p}^\infty, X_{k,t})| \vert \mathcal{F}_{k, t-1,n})\leq \upsilon \alpha^\top|x - y|
 $$
 }
with $\alpha = (\alpha_1, \ldots, \alpha_p).$ Moreover, $E|f(x; n_{k,-1},\ldots,n_{k,-p},\zeta_{k,-1}^\infty, \ldots, \zeta_{k,-p}^\infty, X_{k,0})|<\infty$. Then, from \citet{debaly_truquet_2021} Theorem 4, we obtain the stationary and ergodic solution with $\E|\eta_{k,0}| < \infty, k = 1,\ldots, K$.$\square$

Theorem \ref{th::stability} is a straight consequence of Lemma \ref{lem::generalresult} and follows the Lipschitz property of $x \mapsto \log(\exp(x) + 1 + \delta)$ for any $\delta>0$.  
For the asymptotic results for $\hat\theta_T$, the following assumptions are necessary:

\begin{enumerate}[label = (A.\arabic*)]
\setcounter{enumi}{1}
     \item \label{ass::generalStabilityForTrueParameters} The conditions \ref{ass::generallink} and \ref{ass::integrability} are met, and $\theta_0$ verifies: $ \upsilon \sum_{i = 1}^p |\alpha_{i,0}| < 1.$  
     \item \label{ass::generalMomentForUniformIntegrability} For $k = 1, \ldots, K, \E n_{k,0} < \infty$ and 
    $$
    \E\sup_\theta \left(\frac{\phi_\delta(\eta_{k,0}(\theta_0))}{\phi_\delta(\eta_{k,0}(\theta))}  + |\log \circ \phi_\delta(\eta_{k,0}(\theta))|\right) < \infty.
    $$
    \item \label{ass::generalIdentifiability} For $(\delta, \overline\delta) \in [\delta_-, \infty)^2, (\eta, \overline\eta) \in \R^2$, 
    $$
    \phi_\delta(\eta) = \phi_{\overline\delta}(\overline\eta) \Rightarrow (\delta = \overline \delta, \eta = \overline\eta).
    $$
 \end{enumerate}

\begin{enumerate}[label = (C.\arabic*)]
    \item \label{ass::stabilityForTrueParameters} The conditions \ref{ass::integrability} and \ref{ass::contraction} are met, and $\theta_0$ verifies: $ \sum_{i = 1}^p |\alpha_{i,0}| < 1.$
    \item \label{ass::momentForUniformIntegrability} For $k = 1, \ldots, K, \E n_{k,0} < \infty.$ 
    \item \label{ass::identifiability1} For $k = 1, \ldots, K,$ conditionally  on $X_{k,0}$, the distribution of $\left(\frac{Y_{k,-1}}{n_{k-1}}, \cdots, \frac{Y_{k,-p}}{n_{k-p}}\right)$ is not supported by an hyperplan of $\R^p.$
    \item \label{ass::identifiability2} For $k = 1, \ldots, K,$ the distribution of $X_{k,0}$ is not degenerate.
\end{enumerate}

\begin{lem}
 \label{lem::consistency} Let us suppose that the assumptions \ref{ass::generalStabilityForTrueParameters}--\ref{ass::generalIdentifiability} and \ref{ass::identifiability1}--\ref{ass::identifiability2} are met. Then, almost surely, 
$$
\lim_{T\to\infty} \hat \theta_T = \theta_0.
$$
 \end{lem}

We do not prove Lemma \ref{lem::consistency}. Similar results for time-series models can be found in \citet{diop2021inference}, \citet{aknouche_francq_2020}, and \citet{debaly2021multivariate} among others.
 
\paragraph{Proof of consistency part of Theorem \ref{th::asymptotic}}
We will check \ref{ass::generalStabilityForTrueParameters} to \ref{ass::generalIdentifiability}.
\begin{itemize}
    \item  \ref{ass::generalStabilityForTrueParameters} comes from \ref{ass::stabilityForTrueParameters}. 
    \item One can note that here $\phi_{\delta}(x) = \log(1+\delta + \exp(x))$, and $\phi_{\delta}(x) \geq \log(1+\delta), \phi_{\delta}(x)  \leq \kappa_1(\theta)(1+|x|)$, and $|\log \circ \phi_{\delta}(x) | \leq  \kappa_2(\theta)(1+|x|) + \kappa_3(\theta),$ where $\kappa_i, i =1,2, 3$ are continuous functions of $\theta.$ Then  \ref{ass::generalMomentForUniformIntegrability} holds because $\E\sup_\theta|\eta_{k,0}(\theta)|< \infty.$ Indeed $\E Y_{k,0}/n_{k,0} = \phi_{\delta}(\eta_{k,0}) < \infty$ because $\E|\eta_{k,0}|<\infty.$
    \item For \ref{ass::generalIdentifiability}, we note that
    $$
    \phi_\delta(\eta) = \phi_{\overline\delta}(\overline\eta) \Rightarrow  \delta -\overline \delta = \exp \overline \eta - \exp \eta  ,
    $$
    and $0 = \lim_{\eta\rightarrow-\infty, \overline \eta \rightarrow-\infty}\exp \overline \eta - \exp \eta = \delta -\overline \delta.$ Then $\delta = \overline \delta$, and $\eta = \overline \eta.$$\square$
\end{itemize}

Let us set $\sigma^2_{k,0} = \v\left(\frac{Y_{k,0}}{\lambda_{k,0}}\left\vert\right. \mathcal{F}_{k, -1,n} \vee n_{k,0} \right)$, $\partial_\delta \phi_\delta$ the derivative of $\phi_\delta$ with respect to $\delta$, and $\theta_{-\delta}$ the vector of parameters without $\delta.$ We will consider the following assumptions for the asymptotic distribution of $\hat\theta_T.$
\begin{enumerate}[label = (A.\arabic*)]
\setcounter{enumi}{4}
     \item \label{ass::finitematrix} The function $\phi_\delta$ is twice continuously differentiable and for $k = 1, \ldots, K,$
     {\small
     $$
     \E \frac{\sigma^2_{k,0}}{\phi_\delta^2(\eta_{k,0})}\left[\partial_\delta {\phi_\delta(\eta_{k,0}(\theta_0))}^2 + {\phi'_{\delta}(\eta_{k,0}(\theta_0))}^2\|\nabla_{\theta_{-\delta}}\eta_{k,0}(\theta_0)\|_2^2\right]< \infty , \text{ and } $$ $$   \E \frac{1}{\phi_\delta^2(\eta_{k,0})}\left[\partial_\delta {\phi_\delta(\eta_{k,0}(\theta_0))}^2 + {\phi'_{\delta}(\eta_{k,0}(\theta_0))}^2\|\nabla_{\theta_{-\delta}}\eta_{k,0}(\theta_0)\|_2^2\right]< \infty. 
     $$
    } 
     \item \label{ass::generalInvertible} For $k = 1, \ldots, K,$ the distribution of $(\partial_\delta {\phi_\delta(\eta_{k,0}(\theta_0))}, \phi'_{\delta}(\eta_{k,0}(\theta_0))\nabla_{\theta_{-\delta}}\eta_{k,0}(\theta_0))$ is not degenarate.
  
    \item  \label{ass::generalUniformHessian} For $k = 1, \ldots, K, \E\sup_\theta|W_{k,0}^{i,j}(\theta)| < \infty$, where $W_{k,0}^{i,j}(\theta)$ is one of the following quantities for all pairs $i,j$.
    {\small
    $$
   \frac{1}{\phi_\delta^2(\eta_{k,0}(\theta))}\left(\frac{\phi_\delta(\eta_{k,0}(\theta_0))}{\phi_\delta(\eta_{k,0}(\theta))}+1\right)\dpartial{\phi_\delta(\eta_{k,0}(\theta))}{\theta_i}\dpartial{\phi_\delta(\eta_{k,0}(\theta))}{\theta_j} , 
    $$
    $$
     \frac{1}{\phi_\delta^2(\eta_{k,0}(\theta))}\frac{\phi_\delta(\eta_{k,0}(\theta_0))}{\phi_\delta(\eta_{k,0}(\theta))}\dpartial{\phi_\delta(\eta_{k,0}(\theta))}{\theta_i}\dpartial{\phi_\delta(\eta_{k,0}(\theta))}{\theta_j} ,
    $$
    $$
    \frac{1}{\phi_\delta(\eta_{k,0}(\theta))}\left(\frac{\phi_\delta(\eta_{k,0}(\theta_0))}{\phi_\delta(\eta_{k,0}(\theta))}+1\right)\ddpartial{\phi_\delta(\eta_{k,0}(\theta))}{\theta_i}{\theta_j} 
    .$$
    }
 \end{enumerate}

\begin{enumerate} [label = (AN.\arabic*)]
    \item \label{ass::independencySolution} The $K$ stationary sequences solution of (\ref{eq::modelLogLinearWithoutRandomEffect})--(\ref{eq::lambda}) are independent of each other.
    \item \label{ass::squaredIntegrability1} For $k = 1, \ldots, K,$ $\E n_{k,0}^2 < \infty$ ,
     $$ \E  {\sigma^4_{k,0}} < \infty  .
    $$
   \item \label{ass::uniformhessian} For $k = 1, \ldots, K,$ 
   $$
  \E  |X_{k,0}|_1^4 < \infty , \text{and} \E       {Y_{k,0}^4}  < \infty .
   $$
\end{enumerate}

 \begin{lem}
 \label{lem::generalAN}
 Under the assumptions of Lemma \ref{lem::generalresult}, and if \ref{ass::finitematrix}--\ref{ass::generalUniformHessian} and \ref{ass::independencySolution} hold, then 
 $$\lim_{T\to\infty}  \sqrt{T}(\hat \theta_T - \theta_0) = \mathcal{N}(0, J^{-1}VJ^{-\top}) ,$$
 
where $J = \sum_{k = 1}^K J_k$ and $V = \sum_{k = 1}^K V_k$, 
$$
V_k = \E\left[\frac{1}{\lambda_{k,0}^{2}}\left(n_{k,0} - \frac{Y_{k,0}}{\lambda_{k,0}}\right)^2\dot\lambda_{k,0}\dot\lambda_{k,0}^\top \right], 
J_k = \E\left[n_{k,0}\frac{1}{\lambda_{k,0}^{2}}\dot\lambda_{k,0}\dot\lambda_{k,0}^\top \right], and
$$
$$
\dot\lambda_{k,0} = (\partial_\delta {\phi_\delta(\eta_{k,0}(\theta_0))}, \phi'_{\delta}(\eta_{k,0}(\theta_0))\nabla_{\theta_{-\delta}}\eta_{k,0}(\theta_0))^\top.
$$
 \end{lem}
 
As for Lemma \ref{lem::consistency}, we do not prove Lemma \ref{lem::generalAN}. We refer the interested reader to   \citet{diop2021inference}, \citet{aknouche_francq_2020}, and \citet{debaly2021multivariate}, among others.
 
\paragraph{Proof of asymptotic normality part of Theorem \ref{th::asymptotic}}
For the proof of asymptotic normality part of Theorem \ref{th::asymptotic}, one can note that in the single framework ($k=1$), assumptions \ref{ass::squaredIntegrability1} yield the asymptotic normality of  $\sqrt{T}\nabla \ell_k(\theta_0)$ using the central limit theorem for martingale difference. Next, 

\begin{eqnarray*}
\ddpartial{\ell_{k,t}(\theta)}{\theta_i}{\theta_j} &  = &  \frac{1}{\lambda_{k,t}^2(\theta)}\left(\frac{Y_{k,t}}{\lambda_{k,t}(\theta)}-n_{k,t}\right)\dpartial{\lambda_{k,t}(\theta)}{\theta_i}\dpartial{\lambda_{k,t}(\theta)}{\theta_j}  \\
 & + & \frac{1}{\lambda_{k,t}^2(\theta)}  \frac{Y_{k,t}}{\lambda_{k,t}(\theta)} \dpartial{\lambda_{k,t}(\theta)}{\theta_i}\dpartial{\lambda_{k,t}(\theta)}{\theta_j} \\
 & -  & \frac{1}{\lambda_{k,t}(\theta)} \left(\frac{Y_{k,t}}{\lambda_{k,t}(\theta)}-n_{k,t}\right)\ddpartial{\lambda_{k,t}(\theta)}{\theta_i}{\theta_j} =: I_{k,t}(\theta) + II_{k,t}(\theta) + III_{k,t}(\theta).
\end{eqnarray*}
For the first term,
$$
\sup_\theta |I_{k,t}(\theta)| \leq n_{i,t}\left(\frac{\lambda_{i,t}}{\log(1+\delta\_)}+1\right)\sup_\theta\frac{1}{\lambda^2_{k,t}(\theta)}\dpartial{\lambda_{k,t}(\theta)}{\theta_i}\dpartial{\lambda_{k,t}(\theta)}{\theta_j} ,
$$
and 
$$
\frac{1}{\lambda_{k,t}(\theta)}\dot\lambda_{k,t}(\theta)  \preccurlyeq \kappa_{\delta\_} \left(1,\iota_k, \frac{Y_{k,t-1}}{n_{k,t-1}}\ldots\frac{Y_{k,t-p}}{n_{k,t-p}},X_{k,t}^\top\right)^\top ,
$$
where for $x=(x_1,\ldots, x_d), y=(y_1,\ldots, y_d), x \preccurlyeq y$ means $x_i \leq y_i, i = 1, \ldots, d$, and $\kappa_{\delta\_}$ is a function of $\delta\_.$ Then, $\E \sup_\theta |I_{k,t}(\theta)| < \infty$ under the assumption \ref{ass::uniformhessian}. It can be shown similarly that $\E \sup_\theta |II_{k,t}(\theta)| < \infty$ and $\E \sup_\theta |III_{k,t}(\theta)| < \infty.$ By  the Taylor expansion of $r_T(\cdot)$ between $\hat\theta_T$ and $\theta$,
\begin{eqnarray*}
0 = \sqrt{T}\nabla r_T(\hat\theta_T) &=&   \sum_{k = 1}^K \sqrt{T} \nabla \ell_{k}(\hat\theta_T) \\
 & = & \left(\sum_{k = 1}^K \sqrt{T} \nabla \ell_{k}(\theta_0)\right) + \left(\sum_{k = 1}^K  \nabla^2 \ell_{k}(\theta_0)\right)\sqrt{T}( \hat \theta_T - \theta_0) + o_\P(1) .
\end{eqnarray*}

The independence condition for path  \ref{ass::independencySolution}, assumption \ref{ass::squaredIntegrability1}, and the central limit theorem for martingale difference allows us to conclude $ \sum_{k = 1}^K \sqrt{T} \nabla \ell_{k}(\theta_0)$ converges in distribution to a central Gaussian vector of variance $V$ as $T$ tends to infinity. The assumption \ref{ass::uniformhessian} and ergodic theorem  entail that $\sum_{k = 1}^K  \nabla^2 \ell_{k}(\theta_0)$ converges to $ J$. Moreover, conditions \ref{ass::independencySolution}, \ref{ass::identifiability1}, and \ref{ass::identifiability2} ensure that the  matrix $J$ is invertible.$\square$

\section{Conclusions}
Here we developed a new time-series model to handle data having a time-varying number of sampled individuals. We provided a valid statistical inference procedure and applied the model to assessing the combined effect of climate and SBW outbreak on white spruce tree-ring growth in several sites in eastern Canada. We assumed a fixed number of ecological sites $K$. For future work, we plan to investigate the case of diverging $K$ and the length $n$ of observed series. Because many other ecological studies rely on binary variable or count data, it may be useful to extend the framework of this paper to these data types.

\paragraph*{Acknowledgements}
Funding was provided by the Contrat de service de recherche forestière number 3329-2019-142332177 obtained by PM and MMG from the Ministère des Forêts, de la Faune et des Parcs (Québec, Canada), and a doctoral scholarship from GENES (ENSAE/ENSAI) was obtained by ZMD. We thank G. Tougas for data management and A. Subedi for the map of our study sites.  The authors acknowledge the Quebec Ministry of Forests, Wildlife and Parks (MFFP) and H. Morin for providing data and support for this project.

\paragraph*{Data avaibility statements}
All software and data used in this paper are available at the public zenodo repository \url{https://doi.org/10.5281/zenodo.6340148}.

\paragraph{Author Contributions}
ZMD, PM, and MMG conceived and designed the study. ZMD and PM analysed the data. ZMD wrote the first draft. PM and MMG contributed to the funding and provided statistical input and interpretation. All authors contributed to the revision of the manuscript.

\bibliographystyle{plainnat}
\bibliography{biblio}

\newpage
\section*{Appendix}
 \begin{table}[!htp]\centering
\caption{Estimation results for the quasi-maximum likelihood estimation}\label{tab:simulations}
\scriptsize
\begin{tabular}{lrrrrrrrrrrrrrr}\cline{4-14}
 & & &$\alpha_1$ & $\beta_1$ &$\beta_2$ & $\beta_3$ & $\beta_4$ & $\beta_5$ & $\beta_6$ & $\beta_7$ & $\beta_8$ & $\beta_9$ & $\beta_{10}$ \\\midrule
K &T &Scenario &0.6 &0 &1 &-1 &0.5 &-0.5 &-1.5 & 1.5 &-2 &2 &0 \\ \midrule
5 &50 &1 ~ EQMLE & 0.567 &  -0.018 &  0.985 &  -1.051 &  0.631 &  -0.247 &  -1.565 &  1.445 &  -2.165 &  1.673 &  0.011  \\
& & TSE & 0.060 &   0.086 &  0.185 &   0.286 &  0.178 &   0.309 &   0.332 &  0.219 &   0.298 &  0.211 &  0.104 \\ \cline{3-14}
& &2  ~ EQMLE & 0.517 & 0.019 & 0.862 & -0.772 & 0.580 &  -0.493 &  -1.345 &  1.138 &  -1.709 & 1.908 & -0.015 \\
& & TSE & 0.053 & 0.097 & 0.176 &   0.217 & 0.193 &  0.252 &  0.313 & 0.214 &   0.280 & 0.211 & 0.104 \\ \cline{2-14}
&100 &1  ~ EQMLE & 0.362 & 0.035 &  0.743 & -0.632 &  0.457 &  -0.324 & -1.219 &  1.099 &  -1.655 & 1.521 &  -0.080 \\
& & TSE & 0.065 & 0.101 &  0.169 &   0.303 &  0.218 &   0.379 & 0.401 &  0.219 &  0.333 &  0.232 &  0.093\\ \cline{3-14}
& &2 ~ EQMLE & 0.361 & -0.027 & 0.656 &  -0.678 &  0.294 &  -0.411 & -0.956 & 1.148 & -1.237 &  1.397 &  0.097\\
& & TSE & 0.085 & 0.114 & 0.215 &   0.362 &  0.232 &   0.388 &   0.450 &  0.268 &  0.376 &  0.271 &  0.121  \\ 
\midrule
10 &50 &1  ~ EQMLE & 0.304 &  -0.048 & 0.600 &  -0.487 &  0.476 &  -0.460 &  -0.844 &  1.012 &  -1.048 &  1.311 &  0.023   \\
& & TSE & 0.043 &  0.063 &  0.122 &   0.177 &  0.134 &   0.199 &   0.229  & 0.148 &   0.146 &  0.147 &  0.069    \\ \cline{3-14}
& &2  ~ EQMLE & 0.311 & 0.098 &  0.550 &  -0.547 &  0.243 &  -0.486 &  -0.762 &  1.015 &  -1.233 &  1.061 &  0.061   \\
& &TSE  & 0.046 & 0.084 &  0.153 &   0.220 &  0.175 &  0.258 &  0.256 &  0.183 &  0.235 &  0.174 &  0.079\\ \cline{2-14}
&100 &1  ~ EQMLE & 0.328 &  0.001 &  0.686 &  -0.507 &  0.222 &  -0.293 &  -0.838 &  0.903 &  -1.091 &  1.173 &  0.050 \\
& & TSE & 0.035 & 0.056 &  0.112 &   0.153 &  0.107 &   0.184 &   0.164 &  0.108 &  0.133 &  0.119 &  0.059 \\ \cline{3-14}
& &2  ~ EQMLE & 0.285 & 0.033 &  0.676 &  -0.651 &  0.258 &  -0.072 &  -0.603 &  0.826 &  -1.199 & 1.101 &  0.042 \\
& & TSE & 0.039 &  0.063 &  0.142 &  0.175 &  0.119 &   0.249 &   0.205 &  0.158 &   0.160 & 0.160 &  0.076 \\ 
\midrule
15 &50 &1  ~ EQMLE & 0.546 &  -0.004 &  0.865 &  -0.734 &  0.462 &  -0.467 &  -1.418 &  1.280 &  -1.854 &  1.874 &  0.010 \\
& & TSE & 0.041 & 0.069 &  0.124 &   0.186 &  0.125 &   0.220 &   0.219 &  0.140 &  0.222 &  0.156 &  0.062 \\ \cline{3-14}
& &2  ~ EQMLE & 0.531   0.002 &  0.894 &  -0.985 &  0.405 &  -0.614 &  -1.342 &  1.228 &  -1.845 &  1.692 & 0.0163\\
& & TSE & 0.039 &  0.060 &  0.119 &   0.165 &  0.120 &   0.190 &  0.195 &  0.138 &   0.180 &  0.144 & 0.057 \\ \cline{2-14}
&100 &1  ~ EQMLE & 0.384 & -0.014 &  0.816 &  -0.549 &  0.160 &  -0.546 & -1.226 & 0.987 &  -1.314 &  1.447 & 0.044  \\
& &  TSE & 0.028 & 0.040 & 0.075 &   0.105 &  0.080 &   0.147 &  0.134 &  0.096 &  0.115 & 0.088 &  0.042  \\ \cline{3-14}
& &2  ~ EQMLE & 0.387 & 0.003 &  0.740 &  -0.675 & 0.471 &  -0.356 &  -0.751 &  1.104 &  -1.540 &  1.394 &  0.058  \\
& & TSE & 0.053 &  0.077 &  0.160 &   0.230 & 0.151 &   0.242 &   0.273 &  0.166 &   0.241 &  0.168 &  0.078\\
\midrule
20 &50 &1  ~ EQMLE & 0.370 &  0.018 & 0.613 &  -0.612 &  0.316 &  -0.337 &  -0.787 &  0.915 &  -1.223 &  1.277 &  0.003  \\
& & TSE  & 0.031 &  0.048 & 0.092 &   0.118 &  0.094 &   0.151 &   0.145 & 0.102 &   0.115 &  0.097 &  0.046  \\ \cline{3-14}
& &2  ~ EQMLE  & 0.369 & -0.004 &  0.616 &  -0.745 &  0.350 &  -0.265 &  -1.022 &  0.949 &  -1.266 &  1.282 &  -0.010   \\
& & TSE  & 0.033  & 0.063 &  0.116 &   0.174 &  0.137 &   0.228 &   0.205 &  0.147 &   0.186 &  0.142 &   0.052  \\ \cline{2-14}
&100 &1 ~ EQMLE & 0.339 & 0.002 &  0.534 &  -0.491 &  0.282 &  -0.393 &  -0.802 &  0.862 &  -1.183 &  1.199 &  0.021 \\
& &  TSE & 0.024 & 0.042 &  0.076 &   0.116 &  0.085 &   0.138 &   0.130 &  0.083 &  0.098 &  0.085 &  0.044  \\ \cline{3-14}
& &2 ~ EQMLE & 0.311 & -0.017 &  0.636 &  -0.524 &  0.294 &  -0.173 &  -0.996 &  1.016 &  -1.144 & 1.192 &  0.039 \\
& & TSE & 0.025 &  0.052 &  0.104 &  0.128 &  0.109 &   0.190 &   0.151 &  0.131 &  0.130 & 0.122 &  0.055 \\ \bottomrule
\end{tabular}
\end{table}

 \begin{figure}
       \centering
       \includegraphics[scale = 0.4]{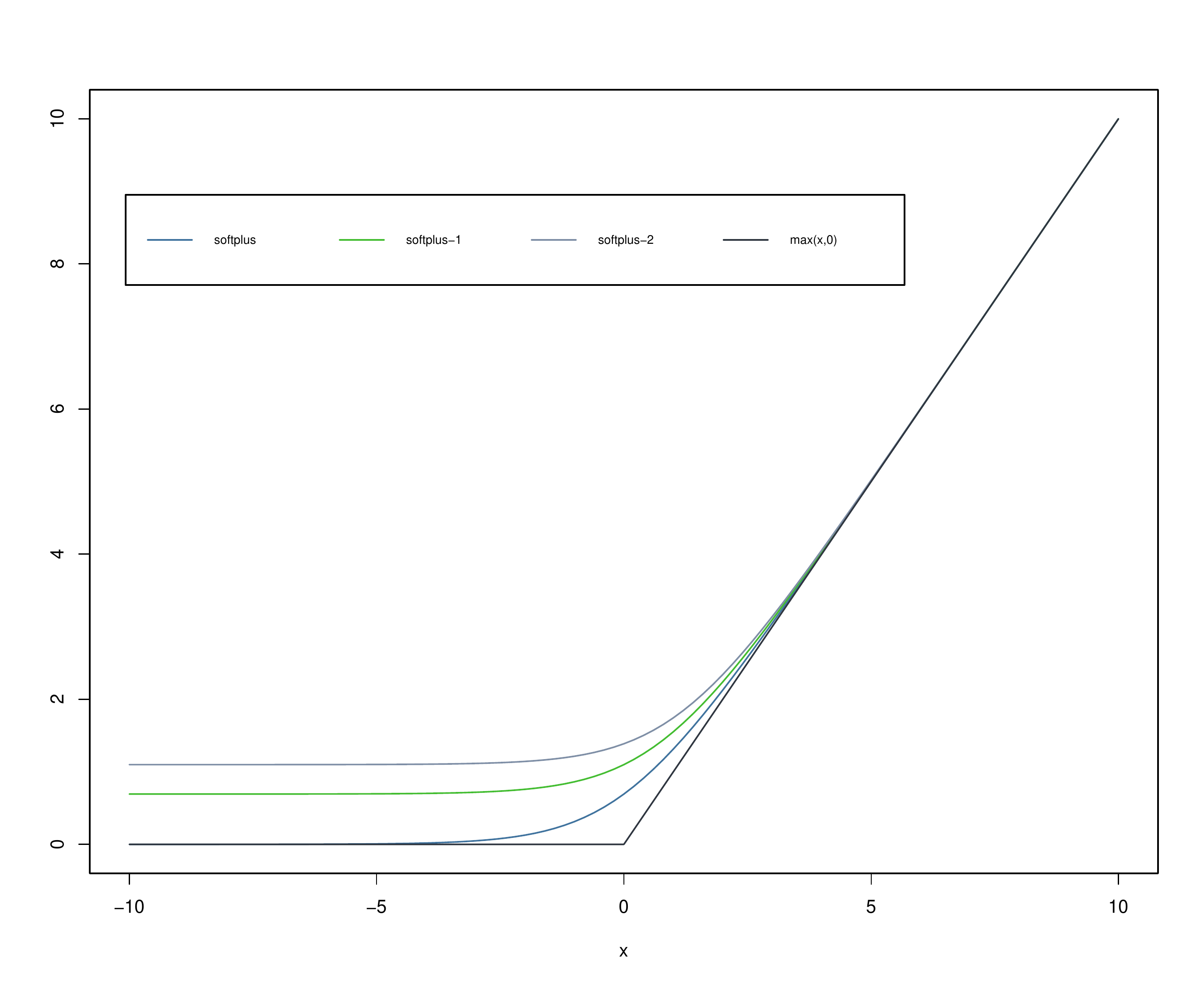}
      \caption{Comparaison between softplus and max(x,0)}
      \label{fig::softplusVsmax}
  \end{figure}
  
  \begin{figure}
      \centering
      \includegraphics[scale = 0.45]{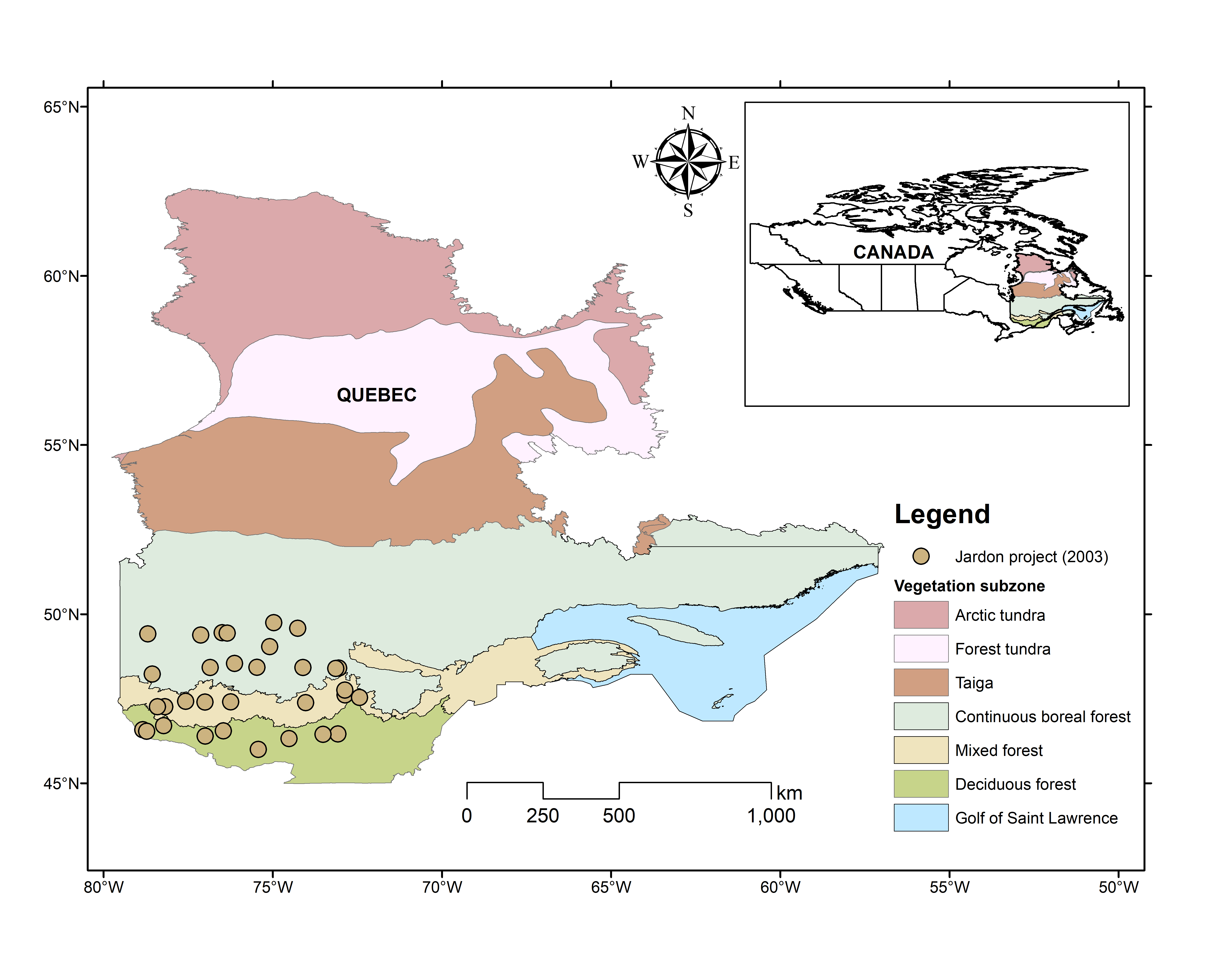}
      \caption{Location of study sites from Jardon Project (2003) in Canadian boreal ecoregions.}
      \label{fig::carteSites}
  \end{figure}
  
\end{document}